# Confinement of Fractional Quantum Number Particles in a Condensed Matter System

Bella Lake[1,2], Alexei M. Tsvelik[3], Susanne Notbohm[1,4], D. Alan Tennant[1,2], Toby G. Perring[5], Manfred Reehuis[1], Chinnathambi Sekar[6,7], Gernot Krabbes[6], Bernd Büchner[6]

[1]*Helmholtz-Zentrum Berlin für Materialen und Energie, Glienicker Strasse 100, 14109, Berlin, Germany.*
[2]*Institut für Festkörperphysik, Technische Universität Berlin, 10623 Berlin, Germany.*
[3]*Department of Condensed Matter Physics and Materials Science, Brookhaven National Laboratory, Upton, New York 11973-5000, United States of America.*
[4]*School of Physics and Astronomy, University of St. Andrews, St. Andrews, Fife KY16 9SS, United Kingdom.*
[5]*ISIS Facility, Rutherford Appleton Laboratory, Chilton, Didcot OX11 OQX, United Kingdom.*
[6]*Leibniz-Institut for Solid State and Material Research, IFW-Dresden, 01171 Dresden, Germany.*
[7]*Department of Physics, Periyar University, Salem-636 011, TN, India.*

**The idea of confinement states that in certain systems constituent particles can be discerned only indirectly being bound by an interaction whose strength increases with increasing particle separation. Though the most famous example is the confinement of quarks to form baryons and mesons in (3+1)-dimensional Quantum Chromodynamics, confinement can also be realized in condensed matter physics systems such as spin-ladders which consist of two spin-1/2 antiferromagnetic chains coupled together by spin exchange interactions. Excitations of individual chains (spinons) carrying spin S=1/2, are confined even by an infinitesimal interchain coupling. The realizations studied so far cannot illustrate this process due to the large strength of their interchain coupling which leaves no energy window for the spinon excitations of individual chains. Here we present neutron scattering experiments for a weakly-coupled ladder material. At high energies the spectral function approaches that of individual chains; at low energies it is dominated by spin 0,1 excitations of strongly-coupled chains.**

The experiments presented in this paper illustrate the condensed matter realization of the confinement idea. The original and most popularized form of this idea comes from particle theory, more specifically from the theory of strong interactions. It is suggested that heavy particles (baryons and mesons) are made of quarks. The latter particles possess properties (more precisely, quantum numbers) which cannot be directly observed, such as fractional electric charge ($\pm 2e/3, \pm e/3$). In a similar fashion in spin ladders excitations of individual spin-1/2 chains (spinons) carry quantum numbers which which are forbidden as soon as the chains are coupled. Quarks are held together by the Yang-Mills (or colour) gauge field which quanta are called gluons. As for any gauge field at smallest distances this interaction obeys the Coulomb law, but at larger distances instead of decreasing it progressively increases due to the gluon-gluon interaction. The popular image is of gluon field lines sticking together and creating some kind of "string" between the quarks. This picture is very appealing since quarks being just end points of a string can under no circumstances appear as individual particles provided the string has a finite tension. Even if one allows the string to snap, none of its pieces can have just one end and hence single quarks still cannot appear. A finite string tension generates an effective potential between the quarks which grows with distance leading to their confinement. Since such a potential well apparently contains infinite number of energy levels, corresponding to different internal energies and hence by the $E = mc^2$ relation to different particles masses, one would imagine that there is an infinite number of particles with the same quantum numbers, but different masses. This picture is oversimplified, however, failing to take into account the quantum nature of the gluon field. Due to the quantum fluctuations the string may snap; as a consequence heavy hadron particles are unstable and decay into lighter ones.

Though the above picture of hadron formation is well established on a qualitative level and has even been extensively popularized[1,2], its quantitative aspects remain unresolved in the sense that it is unknown how to relate the theoretical parameters to the observed hadron masses. This is one of the reasons why condensed matter analogues are interesting since they may provide examples of confinement for which detailed descriptions have been achieved. Numerous analogues exist on the level of models (see, for instance references. 3, 4 and 5); here we report an experimental realization of one particular model namely the weakly coupled spin ladder.

The physics of such a ladder strongly resembles the physics of quark confinement outlined above. The role of quarks in this model is played by excitations of an individual spin chain. These excitations (spinons) carry spin ½. There are several obvious differences. First, the ladder is one-dimensional. Second, the interaction between spinons is not usually attributed to gauge fields, but comes from a short range exchange interactions. While the ladder system we will discuss has isotropic exchange interactions, a simplified picture of confinement resembling the one outlined above holds for chains with a significant Ising-like exchange anisotropy where the ground states have a finite Neel order. In this case spinons can be thought of as domain walls separating two degenerate ground states with opposite staggered magnetization (Fig 1a). Though for a chain with periodic boundary conditions domain walls are always created in pairs, in a single chain there is no energy cost for moving them as far from each other as possible. Hence for a single spin-1/2 chain the spinons do not confine. Since spinons are always produced in pairs, in all experimental probes they appear as incoherent excitations giving rise to energy-momentum continua. That is exactly what our high energy neutron scattering data show. Coupling of the chains leads to spinon confinement. Indeed, as is obvious from Fig 1a, creation of two domain walls at points A and B on a given chain incurs an energy loss proportional to the distance AB-because the reversal of the direction of the spins between A and B costs energy via the interchain exchange interaction with their neighbours on the other chain.

Though such a description gives a good idea of the physics, it omits certain details which become progressively more important when one approaches the point where the exchange interactions are isotropic. The description of confinement where spinons interact by a rigid linear potential holds only if the magnitude of the spinon's spectral gaps well exceeds the inter-chain interaction. This is possible in the Ising limit, but fails when one approaches the point where the exchange is isotropic and the spinons become gapless. Then it becomes impossible to ignore the fact that interaction can create cascades of virtual particles transforming the two-body problem of confinement into a many-body one. The detailed solution taking into account the many-body nature of the confinement was obtained by Shelton *et. al*. in Ref. 6., and a more qualitative discussion can be found in Ref. 7. The result is that unlike for the Ising limit, at the isotropic point confinement of spinons produces only two types particles: the triplet and the singlet excitation branches.

For the confinement physics to be observable one needs to have weakly-coupled spin-ladders. The energy window of weak spinon confinement is $|J_{rung}| \ll E \ll J_{leg}$, where $J_{leg}$ is the intrachain coupling along the chains or 'legs' of the ladder and $J_{rung}$ is the interchain coupled along the 'rungs' of the ladder. Here the dynamical magnetic susceptibility of the ladder has a large spectral weight and at the same time practically coincides with the susceptibility of the individual chains. The region below $J_{rung}$ corresponds to strong confinement. Such weakly-coupled ladders contrast with the more commonly measured strongly-coupled ladder which are always in the strong confinement regime and have only magnon (spin-1) excitation[8]. As we have said, excitations of weakly-coupled spin ladders constitute a triplet of $S = 1$ and a singlet of $S = 0$ particles, which are both located at wavevector transfer parallel to the legs of the ladder of $Q_{leg}=\pi$ and have values of rung wavevector of $Q_{rung}=\pi$ and 0 respectively. For general values of exchange interactions both excitations have spectral gaps and at small interchain coupling these gaps are linear functions of $J_{rung}$ and $J_{cyclic}$, where the latter quantity is the exchange integral for the cyclic (four spin) exchange which can also be found in these systems. The triplet excitation ("triplon") survives for strong rung couplings. No signs of the singlet particle, however, have ever been detected in the strong-coupling limit. Since the theory predicts that for antiferromagnetically (AF) coupled chains its gap increases with the growth of $J_{rung}$, one possibility is that at strong coupling it becomes the two-triplon bound state which is also located at $Q_{rung}=0$[9,10]. Another interesting property of the weak coupling limit is the mirror symmetry between ladders with AF $J_{rung}<0$ and ferromagnetic (F) $J_{rung}>0$ exchange. At $J_{cyclic}=0$ the change of sign of $J_{rung}$ simply leads to interchange of the $Q_{rung}$ locations of the singlet and the triplet excitations. While the rung exchange increases the size of both the singlet and triplet gaps, the cyclic exchange reduces the triplet gap while increasing the singlet gap. Therefore, in the presence of sizable $J_{cyclic}$ one may fine tune the interactions so that the triplet gap vanishes while the singlet gap remains. In this way one reaches a Quantum Critical Point (QCP) associated with a transition from the Haldane spin liquid phase with no local order parameter to a spontaneously dimerized phase which has a local order parameter in the form of staggered energy density[11]. This QCP is described as a particular version of the Wess-Zumino-Novikov-Witten (WZNW)[12,13] model, namely the $SU(2)_2$ WZNW model. Despite being famous among the theorists, only a few condensed matter realizations of this QCP are known. These include the Pfaffian state which was proposed to describe edge states in the $\nu=5/2$ Fractional Quantum Hall effect[14,15], and the spin $S=1$ chain with both conventional and biquadratic exchange interactions[16,17]. As yet however, there has been no experimental observation of this QCP.

In this paper we report neutron scattering data for $CaCu_2O_3$, which is a good candidate for the weakly-coupled ladder. Our observations illustrate some of the points raised above. In particular, they present the first evidence for the existence of the singlet mode. Another curious property of $CaCu_2O_3$ is that the triplet gap in this material happens to be so small[18] that within the experimental accuracy the system appears to be a critical. In the bulk material this critical point is obscured by the phase transition to long-range antiferromagnetic order taking place at $T_N=25$ K. That energy scale, however, is smaller than the energy scales characterizing the individual ladders and therefore does not significantly affect the excitations. Full details of the crystal growth and neutron scattering measurements are given in the methods section.

The band structure calculations predict that in $CaCu_2O_3$ both exchange constants $J_{rung}$, $J_{cyclic}$ are more than an order of magnitude smaller than $J_{leg}$[19,20] (see caption of figure 1 for further details), and the data reported in this paper as well as susceptibility measurements[21,22] support this prediction. In Fig. 2 we show the high energy neutron scattering signal as a function of the wavevector along the ladders. This intensity coincides with the scattering for individual spin $S=1/2$ chains with an intrachain exchange constant of $J_{leg}=-162$ meV, clearly demonstrating that the rung and cyclic couplings are weak in comparison to the bandwidth of the magnetic excitations. Though the rung coupling is smaller than usual, it is still significant since the measurements of the dynamical magnetic susceptibility reveal a strong $Q_{rung}$ dependence below 70 meV as shown in Fig. 3. This makes $CaCu_2O_3$ an ideal system for the theory of confinement developed in references 6 and 11. The band structure calculations also predict a sizable ratio $J_{cycl}/J_{rung}\approx 0.25$[19]. This makes it likely that the triplet gap is smaller than 2.5 meV (the energy equivalent of the Néel temperature) and thus is experimentally unobservable, while the singlet retains a substantial gap.

The rung coupling becomes evident when comparing the magnetic response at different rung wavevectors. Magnetic susceptibility of a periodic array of non-interacting spin-ladders can be written as $\chi = (1+\cos Q_{rung})\chi_b(\omega,Q_{leg})+(1-\cos Q_{rung})\chi_{ab}(\omega,Q_{leg})$, where $\chi_b$ and $\chi_{ab}$ are the bonding and anti-bonding susceptibilities of an individual ladder (for the purposes of this paper we will neglect the inter-ladder interactions assuming that $T > T_N$). The bonding susceptibility which is strongest at $Q_{rung}=0$, is dominated by the simultaneous emission of triplet and singlet modes and the singlet spectrum can be probed by neutron scattering via these triplets. The antibonding susceptibility which is strongest at $Q_{rung}=\pi$ is sensitive only to the triplet mode. Fig. 3 shows that at high energies $J_{rung}\ll E\ll J_{leg}$, $\chi$ is independent of $Q_{rung}$ indicating that there is no difference between bonding and antibonding susceptibilties at these energies and implying that the rung exchange is indeed weak in comparison to the magnetic bandwidth. At the same time at low energies the observed $Q_{rung}$-dependence shows that $\chi_b$ is completely suppressed as would be expected for energies below the singlet gap. This provides clear evidence for both the relevance of the rung and cyclic interactions as well as the existence of the gapped singlet mode.

We briefly comment on the theoretical part of our work. The analytical expressions for the magnetic susceptibility are available only in the continuum limit. To simplify the mathematics we will consider the triplet gap to be zero, which is true within experimental accuracy. Then below $2\Delta_s$, where $\Delta_s$ is the gap in the singlet spectrum, the imaginary part of the anti-bonding susceptibility is dominated by the gapless triplet excitations with linear dispersion $\varepsilon(Q_{leg})=v_s|\pi-Q_{leg}|$ where $v_s$ is the particle velocity ($J_{leg}\pi/2$). In this region it has the form universal for all critical (1+1)-dimensional theories:

$$\chi''_{ab}(\omega,\pi-Q_{leg})=\frac{A}{T^{2-4h}}\operatorname{Im}\left[\rho\left(\frac{\omega-v_s Q_{leg}}{4\pi T}\right)\rho\left(\frac{\omega+v_s Q_{leg}}{4\pi T}\right)\right], \qquad (1)$$

where $\rho(x)=\Gamma(h-ix)/\Gamma(1-h-ix)$ and A is a numerical constant. The hallmark of the WZNW QCP predicted for the weakly-coupled ladder, is the value of the conformal dimension of the staggered magnetization, $h = 3/16$[23]. In contrast the $S=1/2$ Heisenberg AF chain has $h=1/4$[24,25] and is at the Luttinger Liquid QCP. To make a detailed comparison between the theory and experiment we simulate the experiment using a virtual sample whose susceptibility is given by the theory (see methods section). Fig. 4a shows that at high energies the chains are effectively decoupled. The continuum description of single chains works well for the energies where the excitation spectrum is linear which in our case includes energies up to 300 meV. At higher energies we employ the solution of the Bethe ansatz for single chains[26] which takes into account spectral nonlinearities and gives a more accurate description of the data. On the other hand, the low energy data agree best with the prediction for strong confinement. Thus, despite a certain amount of noise present, the antibonding data (blue symbols) displayed on Fig. 4b discriminate between $h=1/4$ (single chain) and $h = 3/16$ (weakly-coupled ladder) in favour of the latter, implying that $CaCu_2O_3$ is close to the WZNW QCP.

The data for the bonding susceptibility (orange symbols in Fig.4b) are also fitted well by the theory, especially at low energies. This part of the susceptibility is not critical, but the analytic expression is still available in terms of correlation functions of the off-critical quantum Ising model[6]. This susceptibility is always incoherent and describes the process of a simultaneous emission of a single singlet excitation with dispersion $\varepsilon_s=\sqrt{\Delta_s^2+[v_s(Q_{leg}-\pi)]^2}$, dressed by gapless triplet excitations. The presence of triplets which unlike the singlet obey the neutron scattering selection rules, make this feature observable with neutrons. Since the corresponding expression for non-zero $T$ is quite complicated, we restrict ourselves to $T=0$, which is sufficient given the limits imposed by the experimental resolution. The derivation of the imaginary part of the dynamical bonding susceptibility is outlined in the methods section and is given by

$$\chi''_b\left(\omega,\pi-Q_{leg};T=0\right)=D\theta(s^2-1)s^{-4h}(s^2-1)^{4h-1}F(2h,2h,4h;1-s^{-2}) \qquad (2),$$

where $s^2 = (\omega^2 - v_s^2 Q^2)/\Delta_s^2$, $F(a,b,c;x)$ is the hypergeometric function, D is a numerical coefficient and $h = 3/16$. Eq.2 is the result of convolution of the singlet mode propagator with the critical correlation function of the triplet modes. The fit shown in Fig. 4b yields a singlet gap of $\Delta_s$ ~16meV.

We conclude with a qualitative description of the QCP and a few proposals for the further work. Experiments which may give further support for our suggestion that $CaCu_2O_3$ is close to the $SU(2)_2$ WZNW QCP include measurements of the low temperature specific heat and the Knight shift. The physical picture of this QCP is quite simple since it turns out that it can be described as a theory of weakly interacting Majorana fermions. Unlike the more familiar Dirac fermions, the Majorana ones do not have antiparticles and hence have the same spectrum as phonons ($\varepsilon(Q_{leg}) = v_s |\pi - Q_{leg}|$). As a consequence, one Majorana mode occupies ½ of the Hilbert space of a conventional fermion. At energies much smaller than the singlet gap they occupy ¾ of the Hilbert space of the pair of spin S=1/2 chains that form the ladder. Indeed, adding two spin ½ states produces one singlet state and one triplet state which is three times degenerate. The former state has a gap and therefore does not appear at low energies. This simple description has consequences for the thermal properties. At $T_N \ll T \ll \Delta_s/k_B$ the magnetic contribution to the specific heat per unit length of a single ladder must be $C_v = \pi T C / 3 v_s$, where $C = 3/2$ is the central charge characterizing the above QCP. This value of $C$ reflects the fact that the gapless triplons occupy only ¾ of the Hilbert space of non-interacting chains so that the ladder with both $J_{rung}$, $J_{cycl}$ switched off must have C=2. Another check for the validity of our interpretation of the low energy properties can be done by NMR measurements. The scaling dimension $h$ can be independently extracted by measuring the Knight shift on $Cu^{2+}$ ions. The theory predicts[6] $K \sim \int dQ_{leg} \chi(\omega=0, Q_{leg}) \sim T^{4h-1}$, where $h = 3/16$ well below the singlet gap and $h = 1/4$ well above it. We also intend to repeat our neutron measurements for several temperatures to establish the energy/temperature critical scaling characteristic of the WZNW QCP.

## Methods

### 1. Derivation of Eq. 2

As was shown by Shelton et. al.[6], the continuum limit of spin ladder can be well described by the theory of four non-interacting Majorana fermions or, equivalently, the theory of four quantum Ising (QI) models. The most transparent way to describe a QI model is to define it on a lattice (though at the end we need to consider the continuum limit). Then the Hamiltonian is

$$H = -\sum_n [J \sigma_n^z \sigma_{n+1}^z + (J + \Delta) \sigma_n^x] \quad (3)$$

Where $\sigma^z$, $\sigma^x$ are Pauli matrices. The spectrum of this model is

$$\varepsilon(k) = \sqrt{\Delta^2 + 4J(J+\Delta)\sin^2(k/2)} \quad (4)$$

and in the continuum limit $k \ll 1$, $\Delta \ll J$ becomes $\varepsilon(k) \approx \sqrt{\Delta^2 + (vk)^2}$. In our case $J = \pi J l_{eg}/2$ so that the spectrum would match the one of an individual chain at high energies, and $m$ is related to the inter-chain exchange. More precisely, three of QI's have the same gap $\Delta = \Delta_t = A J_{rung} - B J_{cyclic}$ and the fourth one has $\Delta_s \sim -(A J_{rung} + 3 B J_{cyclic})$, where A,B are some numerical coefficients. Though the sign of $\Delta$ does not affect the spectrum, it has a crucial effect on the ground state and hence the behaviour of the correlation functions of model (3). Indeed, at $\Delta > 0$ the average $<\sigma^z> = 0$ in the ground state and at $\Delta < 0$ this average is finite.

Model (3) possesses a remarkable property of self-duality, namely, one can introduce new set of Pauli matrices $\mu^a_{n+1/2}$ ($a = x,y,z$) defined on the dual lattice such that

$$\mu^z_{n+1/2} = \prod_{j=1}^{n} \sigma^x_j, \quad \mu^x_{n+1/2} = \sigma^z_n \sigma^z_{n+1} \tag{5}$$

So that the Hamiltonian (3) becomes $H = -\sum_n [(J+\Delta)\mu^z_{n-1/2}\mu^z_{n+1/2} + J\mu^x_{n+1/2}]$. Since $\mu^z$ and $\sigma^z$ cannot have nonzero ground state values simultaneously, at $\Delta < 0$ when $<\sigma^z> \neq 0$ one has $<\mu^z> = 0$ and at $\Delta > 0$ one has $<\mu^z> \neq 0$. It is customary to call $\sigma = \sigma^z$ the order and $\mu = \mu^z$ the disorder parameter fields respectively.

According to Shelton $et.al.$[6], the spin correlation functions of the spin-1/2 ladder can be expressed in terms of correlation functions of the QI models. Namely, we the following relationships hold:

$$\vec{n}_{ab} = (\mu_1\sigma_2\sigma_3, \sigma_1\mu_2\sigma_3, \sigma_1\sigma_2\mu_3)\mu_0 \tag{6}$$

$$\vec{n}_b = (\sigma_1\mu_2\mu_3, \mu_1\sigma_2\mu_3, \mu_1\mu_2\sigma_3)\sigma_0 \tag{7}$$

for antibonding and bonding components of staggered magnetization. Since the correlation functions of these operators are known, one can calculate correlation functions of staggered magnetizations of the ladder. In the situation relevant to $CaCu_2O_3$ we have $\Delta_t = 0, \Delta_s < 0$ so that the triplet sector is critical and the singlet sector is in the disordered state. At the critical point one has

$$<<\sigma(\tau,x)\sigma(0,0)>> = <<\mu(\tau,x)\mu(0,0)>> \sim [\tau^2 + (x/v)^2]^{1/8} \tag{8}$$

At energies smaller than $2\Delta_s$ one can replace $\mu_0$ by its average value and in the correlation function of $\sigma_0$ to leave only the part corresponding to the emission of one particle:

$$<<\sigma_0(\tau,x)\sigma_0(0,0)>> \sim K_0(\Delta_s\sqrt{\tau^2 + (x/v)^2}) \tag{9}$$

Substituting (8,9) into the correlation function of the staggered fields given by Eqs.(6,7) and taking the Fourier transform we arrive to Eq.(2) of the main text.

## 2. Single Crystal Growth

Single crystals of $CaCu_2O_3$ were grown using the traveling solvent floating zone method with CuO as flux[27]. Initially powder samples of $CaCu_2O_3$ were sintered at 1020°C for six days in oxygen atmosphere with intermittent grindings. As $CaCu_2O_3$ is a thermodynamically unstable phase below 1000°C, the samples were quenched at the end of every run. Feed rods for crystal growth (6 mm diameter 10 cm long) were made, pressed hydrostatically under 15 kN/cm$^2$ and further densified at 1020°C for 75 hours. Crystal growth was carried out in an infrared radiation furnace equipped with four ellipsoidal mirrors (Crystal Systems Inc.) and 300-W halogen lamps were used to obtain a steep temperature gradient. Previous work suggests that the CuO poor starting charge (70% CaO and 30% CuO) is the optimum concentration of the solvent rod. About 0.5 g of this flux rod was fixed at the end of the feed rod. Stable growth was achieved for a pulling rate of 1.0 mm/h, rotation rate of 30 rpm and 5 bar oxygen pressure. At the end of the run the power supply was reduced rapidly. A large portion of the boule was found to be inclusion free, although the peripheral region contained a thin layer of the impurity phase $Ca_2CuO_3$ which was removed by cutting.

## 3. Experiments

The neutron scattering data were collected using the MAPS time-of-flight spectrometer at the ISIS neutron spallation source in the Rutherford Appleton Laboratory, U.K. The $CaCu_2O_3$ sample consisted of several co-aligned single crystal pieces with total weights of 13.2g and 8.9g for the two experiments that were performed. Both the legs of the ladder (*b* axis) and the rungs of the ladder (*a* axis) were perpendicular to the incident neutron beam with the legs being horizontal and the rungs vertical. This orientation allowed the sample to be probed along both the rung and leg directions. A Fermi chopper was phased to select neutrons of a fixed incident energies ranging from 30meV to 1000meV. The resolution was controlled by the chopper speed and frequencies ranging from 150Hz to 500Hz were used. The sample was cooled in a closed cycle cryostat and data was collected at a few temperatures from 9K to 300K. Counting times for each setting were typically between 1200 to 8600μAmp (7-50 hours). The data was corrected for detector efficiency and normalized to absolute units by measuring the incoherent neutron cross-section of a vanadium sample of known mass in a white beam as well as in monochromatic beams corresponding to each setting of energy and chopper for which data was collected. The non-magnetic intensity arising from the coherent and incoherent multi-phonon scattering from the sample and its environment was subtracted. This was achieved by smoothing the data in non-magnetic regions of reciprocal space (e.g. $Q_{leg}$=0,1,2, etc) and interpolating it to estimate the background at regions where magnetic signal is present. In spite of the long counting times and the careful data treatment some noise remained in the data due to the weak magnetic signal. Additional measurements were performed using a sample orientation where the rungs of the ladder were parallel to the incident beam while the legs of the ladder were perpendicular and horizontal. These measurements were used to probe the interladder dispersion along the *c* direction.

## 4. Simulations

The data was compared to several theories. This was achieved by simulating the experiment with a virtual sample whose $\chi''(Q,E)$ is that given by the theory using the ms_simulate program in the mslice package. The theoretical $\chi''(Q,E)$ was converted to $S(Q,E)$ by multiplying by the thermal occupation factor and was also corrected for the anisotropic form factor of copper, the mosaic spread of the sample and the energy resolution. In the simulation the model sample had the same orientation as in the real $CaCu_2O_3$ experiment and the same configurations were also used (incident energy and chopper speed). All the same manipulations performed on the real data were also performed in the virtual data (except normalization and background subtraction). The data could then be compared to theory by performing identical cuts and slices as shown in figure 4.

**Acknowledgements**
We are grateful to A. A. Nersesyan, F. H. L. Essler, J.-S. Caux, R. Coldea and T. M. Rice for interesting discussions. E.M. Wheeler helped with the initial data analysis. The work was supported by the US DOE under contract number DE-AC02-98 CH 10886 (AMT). AMT also thanks Galileo Galilei Institute for Theoretical Physics for kind hospitality and INFN for partial support during the completion of this work.

**Author Contributions**
C.S., G.K. and B.B. grew the crystals. B.L., S.N., D.A.T, T.G.P. and M.R. did the experiments. Data analysis was performed by B.L., theory was done by A.M.T. and the paper was written by A.M.T. and B.L.

**Competing Financial Interests Statement**
The authors declare that they have no competing financial interests.

**Correspondence**
Correspondence and requests for materials should be addressed to B.L. (bella.lake@helmholtz-berlin.de) and A.M.T. (tsvelik@cmt9.phy.bnl.gov)

**Figures**

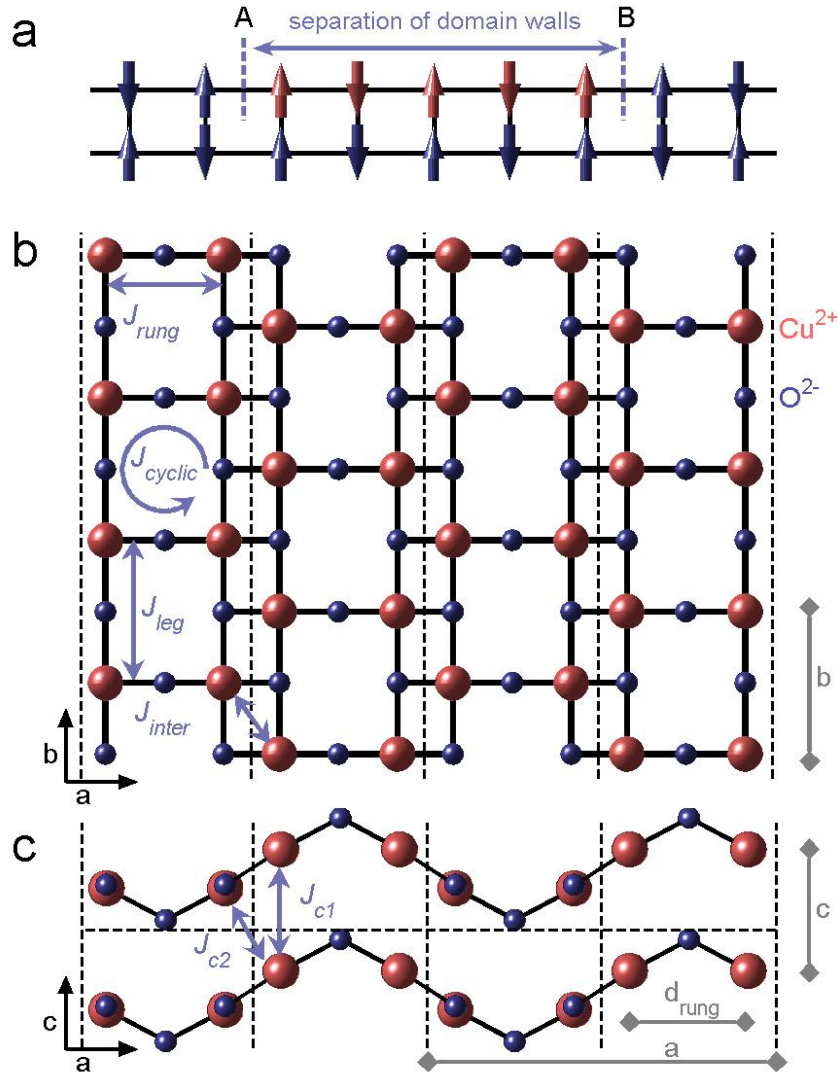

**Figure 1** - An illustration of confinement on a ladder and the structure and magnetic interactions of $CaCu_2O_3$. **a)** shows how the region between two spinons (domain walls) on a chains consists of reversed spins (coloured in red); if this chain is coupled antiferromagnetically to another chain as in a spin ladder these reversed spins cost energy due to their parallel alignment with the spins on the neighbouring chain. This energy cost which is proportional to the separation of the spinons acts to confine the spinons. **b)** and **c)** illustrate the structure of $CaCu_2O_3$ for the *a-b* plane and *a-c* plane respectively. $CaCu_2O_3$ has orthorhombic symmetry with space group *Pmmn* and lattice parameters *a*=9.949 Å, *b*=4.078 Å, and *c*=3.460 Å at *T*=10K. The magnetic $Cu^{2+}$ ions have spin=1/2 and are represented by the red symbols, they are coupled to each other by superexchange interactions via the $O^{2-}$ ions (blue symbols) and the Cu-O bonds are represented by the solid black lines; the $Ca^{2+}$ ions are not shown. The lattice parameters are shown in grey as well as the rung distance $d_{rung}$ which is approximately one third of the *a* lattice parameter. The structure consists of copper oxide layers stacked along the *c* direction, the ladders lie within this plane running parallel to *b* and neighbouring ladders are shifted by half a unit cell in *a*. The dotted black lines indicate the separate ladder units and the inter- and intraladder exchange interactions are labeled. The coupling along the legs, $J_{leg}$, occurs via superexchange interactions mediated by oxygen, the Cu-O-Cu bond angle is 180° giving rise to strong antiferromagnetic coupling (according to the Goodenough-Kanamori-Anderson rules). In contrast the Cu-O-Cu

bond along the rungs is 123° and therefore $J_{rung}$ is expected to be substantially weaker although still antiferromagnetic. In addition a weak antiferromagnetic interaction, $J_{diag}$, is predicted between opposite copper ions within each plaquette of the ladder. The ladders are coupled together by a number of weaker interactions. Within the ***a-b*** plane, $Cu^{2+}$ ions on neighbouring ladders are connected via Cu-O-Cu bonds that are 90° giving rise to a weak ferromagnetic $J_{inter}$. Note that $J_{inter}$ is frustrated and competes with the much stronger $J_{leg}$, thus its energy cancels in the Hamiltonian to first order. Weak interladder couplings $J_{c1}$ and $J_{c2}$, are also expected between ladders in the ***c*** direction. Finally, in common with other planar copper oxide materials, $CaCu_2O_3$ is expected to have a four spin cyclic exchange interaction, $J_{cyclic}$, coupling the four copper ions that form each plaquette. Quantum chemistry calculations give the following exchange constants for $CaCu_2O_3$ $J_{leg}$=-147 to -134 meV; $J_{rung}$=-15 to -11.3meV; $J_{cyclic}$=4meV; $J_{inter}$<24meV; $J_{diag}$=-0.2meV; $J_{c1}$=0.1meV; $J_{c2}$=0.8meV[19,20]. Susceptibility data fitted to a spin-1/2 Heisenberg chain model without other interactions provide good agreement with the data and suggest that $J_{leg}$ is indeed the dominant interaction and has a value of 168meV[22].

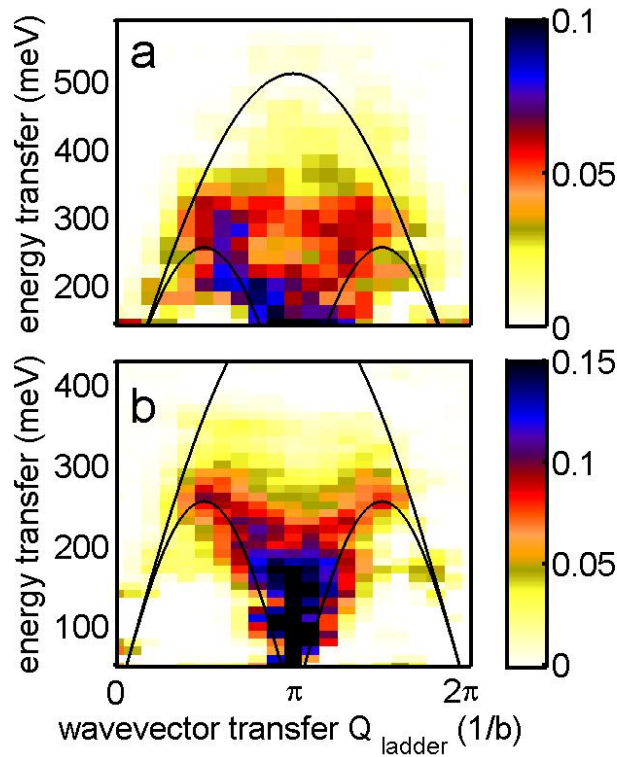

**Figure 2** - High energy inelastic neutron scattering data for $CaCu_2O_3$. The data is displayed as a function of energy (*E*) and wavevector parallel to the ladder direction ($Q_{ladder}$), and is integrated over all wavevectors perpendicular to the ladder. The non-magnetic background has been subtracted (see methods section) and the colours give the strength of the magnetic neutron scattering cross-section, *S*(*Q*, *E*). The solid black lines indicate the upper and lower boundaries of the multi-spinon continuum of a spin-1/2 Heisenberg antiferromagnetic chain with $J_{leg}$=162meV. The data was collected using the MAPS time-of-flight spectrometer at ISIS, Rutherford Appleton Laboratory, U.K. (methods section). The measurements were performed above the Néel temperature at 35K. Two different settings of incident energy and chopper speed were used: **a)**, 1000meV, 400Hz to give an overview of the continuum and **b)**, 600meV, 500Hz to provide more detail of the lower energy region.

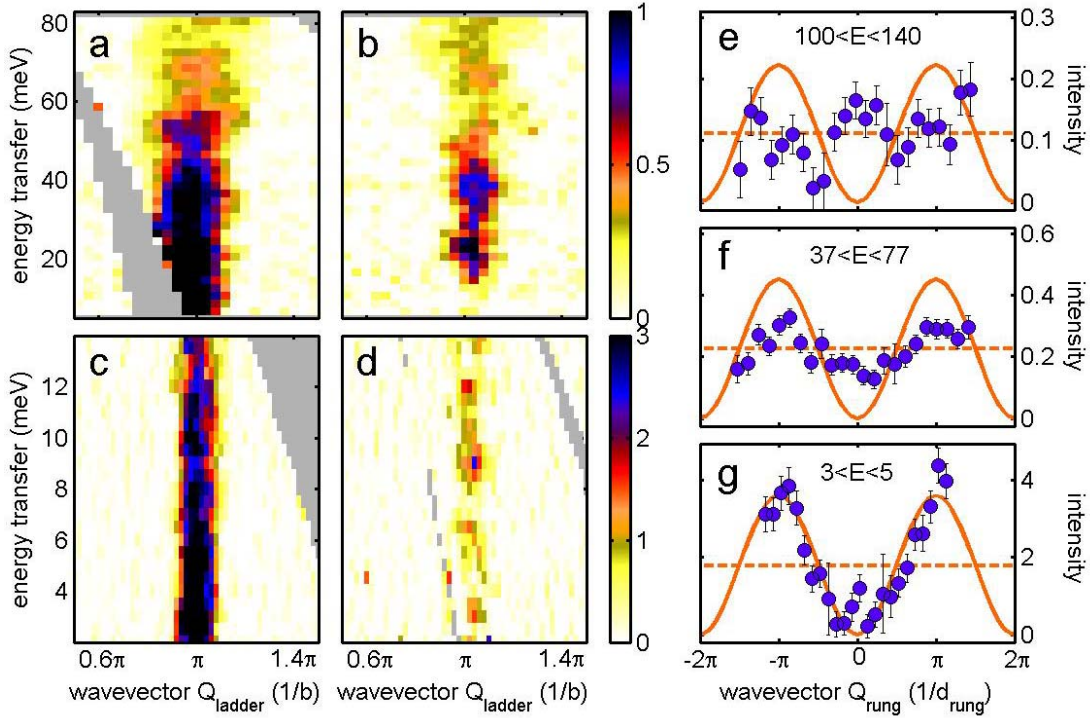

**Figure 3** - Low energy inelastic neutron scattering data for $CaCu_2O_3$. Panels **a)**, **b)**, **c)**, and **d)**, shows the background subtracted $S(Q, E)$ displayed as a function of energy ($E$) and wavevector parallel to the ladder direction ($Q_{ladder}$). The colours indicate the intensity and the grey shaded regions give the areas without detector coverage. The data is integrated over all wavevectors perpendicular to the copper oxide planes ($Q_c$). **a)** and **c)** show the antibonding susceptibility and are integrated over wavevectors parallel to the rung in the range $0.74\pi < Q_{rung} < 1.26\pi$ where $Q_{rung}$ is expressed in units of $1/d_{rung}$ and $d_{rung}$ is the rung distance (see figure 1). **b)** and **d)** show the bonding susceptibility and are integrated over the range $-0.26\pi < Q_{rung} < 0.26\pi$. **e)**, **f)**, and **g)** show the background subtracted $S(Q, E)$ for $0.8\pi < Q_{ladder} < 1.2\pi$ and all $Q_c$, integrated over the energy ranges 100meV<$E$<140meV, 37meV<$E$<77meV and 3meV<$E$<5meV respectively and plotted as a function of $Q_{rung}$. The errorbars represent standard deviations given by $N^{1/2}$ and normalised for the total proton charge. The solid orange line gives the expected modulation of the antibonding susceptibility which is of the form $(1-\cos(Q_{rung}))$. This modulation is observed in the data for 3meV<$E$<5meV (panel **g**) suggesting that this energy is below the gap in the bonding susceptibility. Note that we are not simply observing the interladder dispersion here because that would have a periodicity three times smaller of the form $|\sin(Q_a)|\approx|\sin(3Q_{rung})|$ due to the fact that the rung distance is 1/3 of the $a$ lattice parameter ($d_{rung}\approx a/3$) (see Fig. 1). In fact the lack of a modulation with this periodicity suggests that the dispersion of the triplet in this direction has a maximum energy less than 3meV. Other measurements (not given here) reveal that the dispersion in the $Q_c$ direction has a maximum energy of 5meV showing that the interplanar coupling is very weak. The measurements were performed at 35K with incident energy and chopper settings of **a)**, **b)** and **f)** 102meV 300Hz; **c), d)** and **g)** 30.5meV 200Hz; **e)** 162meV 200Hz; respectively.

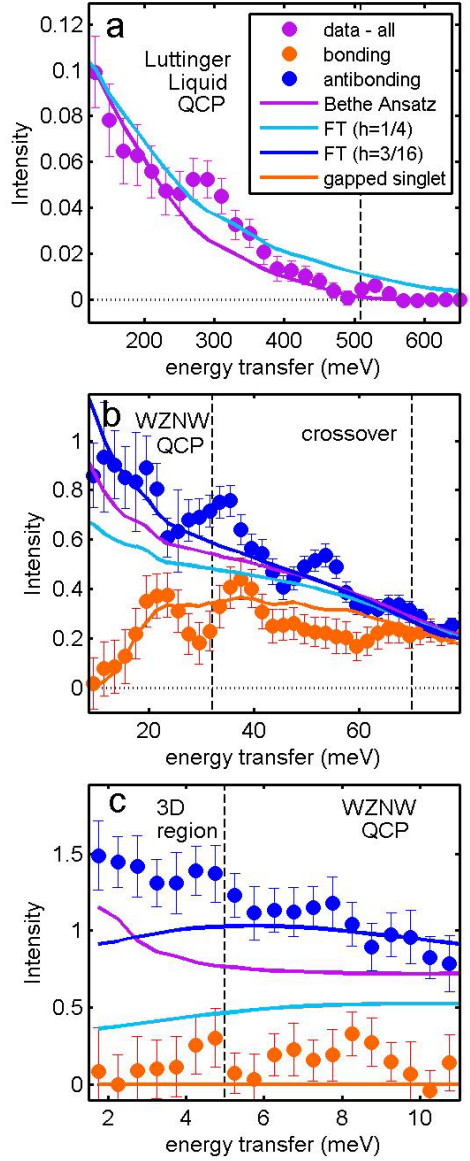

**Figure 4** – Comparison between data and theory. Panel **a**) shows the background subtracted $S(Q, E)$ displayed as a function of energy for high energies. The data is summed over wavevector parallel to the ladder in the range $0.8\pi < Q_{ladder} < 1.2\pi$ and is integrated over all wavevectors perpendicular to the ladder both $Q_c$ and $Q_{rung}$. The data are compared to the nearest neighbour Heisenberg antiferromagnet, two solutions of this model were simulated; the solution of the Bethe Ansatz[26] and the field theory (FT) expression for a Luttinger liquid QCP[24] (equation (1) with $h=1/4$). Details of the simulation are given in the methods section. Panels **b**) and **c**) show the low energy background subtracted $S(Q, E)$ summed over $0.8\pi < Q_{ladder} < 1.2\pi$ and integrated over all $Q_c$. The red data points give the bonding susceptibility (integration range $-0.26\pi <Q_{rung}< 0.26\pi$) while the blue data points give the antibonding susceptibility ($0.74\pi <Q_{rung}< 1.26\pi$). The antibonding data is compared to the field theory expressions for the Luttinger liquid[24] and WNZW[23] QCPs (equation (1) with $h=3/16$) as well as the Bethe Ansatz[26]. The bonding data is compared to the expression for the singlet mode with a gap of $\Delta_s=16$meV (equation (2)). The various regimes of the ladder are labeled and their energy ranges are indicated by the vertical dashed lines. The measurements were performed at $T=35$K with incident energy and chopper settings of **a**) and 1000meV, 400Hz; **b**) 102meV, 300Hz; **c**) 30.5meV, 200Hz. The errorbars represent standard deviations given by $N^{1/2}$ and normalised for the total proton charge.